\documentclass{article}

\usepackage{arxiv}

\usepackage[utf8]{inputenc} 
\usepackage[T1]{fontenc}    
\usepackage{hyperref}       
\usepackage{url}            
\usepackage{booktabs}       
\usepackage{nicefrac}       
\usepackage{microtype}      
\usepackage{lipsum}
\usepackage{graphicx}
\usepackage{amsmath,amssymb,amsfonts}
\usepackage{algorithmic}
\usepackage{comment}
\usepackage{nomencl}
\usepackage{enumitem}   
\graphicspath{ {./images/} }

\title{CD-SEIZ: Cognition-Driven SEIZ Compartmental Model for the Prediction of Information Cascades on Twitter}

\author{
    Ece Çiğdem Mutlu \\
  Department of Industrial Engineering and Management Systems\\
  University of Central Florida\\
  \texttt{ece.mutlu@ucf.edu} \\
   \And
Amirarsalan Rajabi\\
  Department of Computer Science\\
  University of Central Florida\\
  \texttt{amirarsalan@knights.ucf.edu} \\
  \And
Ivan Garibay\thanks{Corresponding author} \\
  Department of Industrial Engineering and Management Systems\\
  Department of Computer Science\\
  University of Central Florida\\
  \texttt{igaribay@ucf.edu} \\
}

\begin{document}
\maketitle
\begin{abstract}
Information spreading social media platforms has become ubiquitous in our lives due to viral information propagation regardless of its veracity. Some information cascades turn out to be viral since they circulated rapidly on the Internet. The uncontrollable virality of manipulated or disorientated true information (fake news) might be quite harmful, while the spread of the true news is advantageous, especially in emergencies. We tackle the problem of predicting information cascades by presenting a novel variant of SEIZ (Susceptible/ Exposed/ Infected/ Skeptics) model that outperforms the original version by taking into account the cognitive processing depth of users. We define an information cascade as the set of social media users' reactions to the original content which requires at least minimal physical and cognitive effort; therefore, we considered retweet/ reply/ quote (mention) activities and tested our framework on the Syrian White Helmets Twitter data set from April 1st, 2018 to April 30th, 2019. In the prediction of cascade pattern via traditional compartmental models, all the activities are grouped, and their summation is taken into account; however, transition rates between compartments should vary according to the activity type since their requirements of physical and cognitive efforts are not same. Based on this assumption, we design a cognition-driven SEIZ (CD-SEIZ) model in the prediction of information cascades on Twitter. We tested SIS, SEIZ, and CD-SEIZ models on 1000 Twitter cascades and found that CD-SEIZ has a significantly low fitting error and provides a statistically more accurate estimation.   
\end{abstract}


\section{Introduction}
\label{sec:1}
The widespread use of online social networks (OSNs) for the free sharing of ideas, combined with an ability to reach multiple people simultaneously, and the ease of participation in many communities made OSNs a fundamental source of human interactions-based research studies. Recently, there has been an interest in the literature that try to understand community detection \cite{li2019opinion}, consensus formation \cite{lamiran2019identifying}, link prediction \cite{ahmed2016supervised}, polarization in political views \cite{garimella2017long} etc. OSNs not only offer a huge amount of data every day but also provide a detailed dynamic of human interactions in diverse contexts. Researchers from many disciplines employed these fast-paced datasets to predict the trending topics \cite{xie2016topicsketch}, sentiments of user-generated contents \cite{zhang2016sentimental}; to understand self-organization of users \cite{khazraee2018twitter}, network's influence on sharing behaviour of online audiences \cite{garcia2017understanding}; or to develop recommendation system based on activities \cite{belhadi2020data} on OSNs. The most interesting applications of OSNs are the studies wherein the dynamics of information formation, dissemination, and manipulation are analyzed. This diffusion of information yields fast updates on current events for users from different countries and might be very beneficial in emergencies such as natural disasters. On the other hand, some malicious users may try to manipulate or disorientate the true information and causes quite harmful situations by spreading false information through these tools. Therefore, analyzing the spread of information diffusion helps us to understand its dynamics, use it for useful purposes, and even mitigate the propagation of false information on OSNs. One of the most challenging tasks among information diffusion research is understanding the cascading behavior of user-generated contents. A very detailed taxonomy of the methods used to understand the popularity of a content on OSNs is given in Gao et al.'s study \cite{gao2019taxonomy}. These prediction methods can be classified into three main categories: feature-based (ML) methods, time series-based (stochastic process, deterministic compartmental) methods, and collaborative filtering (matrix factorization, tensor decomposition, transfer learning) methods. In this study, we employ and improve the SEIZ compartmental model to predict the cascade size prediction and information spread. 

Compartmental methods are mathematical approaches commonly used in epidemiology to model the contagious disease transmission. In these models, individuals within a target population are classified into mutually exclusive compartments based on their current status, and their future status is predicted with the specific rate of contact between different compartments and their certain transition rates. The most simple example of the SI model, for example, divides the whole population into two compartments called \textit{Susceptible} (S) agents who are not infected yet and \textit{Infected} (I) agents who are infected by another infected agent. Although this model gives preliminary insights into the transition of a disease, its simplistic assumptions can provide a better understanding of contagion dynamics. Therefore, the most comprehensive version called the SIS model is used to model the dynamics of the spread of disease. In this model, the agents in different compartments can transition back and forth between \textit{Susceptible} and \textit{Infected} statuses. This model enables defining the transition between states continuous and, therefore, quite powerful in modeling diseases such as flu, cold, and allergies. Another more complex SI model version is known as the SIR model in which individuals can \textit{Recover} after a certain period being  \textit{Infected}. Although this model is efficient among epidemic models, its application is limited to predict the popularity of information since the "recovering" is not a transferable status for OSN context. The SEIZ model proposed by Bettencourt et al. \cite{bettencourt2006power}, on the other hand, introduces an \textit{Exposed} (E) and \textit{Skeptic} (Z) states in addition to  \textit{Susceptible} (S) and \textit{Infected} (I) statuses which yields better modeling of information diffusion phenomena using compartmental models. Although this pioneering study aims to model the adoption of Feynman diagrams, later studies demonstrated the superiority of this cascade prediction method on OSNs \cite{bettencourt2006power,jin2013epidemiological,isea2017new,tipsri2015effect,jin2014misinformation}.

In the analysis of user-generated contents' cascading behaviors, the definition of cascade plays a crucial role. On the contrary of the studies focuses on retweet cascades only in which information propagates only via \textit{Retweet} action \cite{riquelme2016measuring,jin2014misinformation,bettencourt2006power}, many studies demonstrated that \textit{Reply} and \textit{Quote} activities play a major role in the rise of the original tweets' virality and improve the popularity of them. Here, we define cascade as the set of reactions to the original content which requires at least minimal physical and cognitive effort; therefore, we considered all \textit{Retweet}, \textit{Reply} and \textit{Quote} activities as a considerable reactions as in \cite{kessling2019analysis,fink2016complex,levens2019using}. In this study, we aim to employ the SEIZ model and improve it considering cognitive factors for the use of Twitter cascades specifically. 

Our contributions are as follows:
\begin{itemize}
    \item We are the first study that employs compartmental models in predicting the size of information cascades on Twitter by using not only \textit{Retweet} but also \textit{Reply} and \textit{Quote} activities which are proved to have a significant effect in the diffusion of information.
    
    \item In addition to employing SEIZ model, we also improved its performance by defining varying transition rates between compartments for each type of activity on Twitter. 
    
    \item Although studies that demonstrate the SEIZ model's superior performance used a minimal number of information cascades (up to 8 at maximum), we show the comparison between different compartmental models on 1000 different cascades, for the first time, to prove the scalability and robustness of our results. Due to the same reason, we are the first study, as far as we know, to demonstrate the superiority of the SEIZ model and our proposed framework statistically. 
\end{itemize}

\section{Methodology}
In this study, we employ commonly-used compartmental models of SIS and a recently-popular model of SEIZ to model the information propagation on Twitter. As aforementioned, SEIZ model demonstrated superior performances in the modeling the information cascades on online social networks; however, researchers employed this model to predict only retweet behaviour based on the assumption that information propagation can be described with the activity of re-sharing (retweet in Twitter) of the specific user-generated content. On the other hand, information can diffuse in a variety of ways on Twitter. \textit{Retweet} is one of the most common diffusion tools in which users have shared the content as is. Furthermore, users can \textit{Reply} the original content and share their comment on it, which may trigger the exposition of the original and reply content of other users at the same time. \textit{Quote} is also considered as a type of \textit{Retweet}, which enables the re-share of the original content by adding your own comment on it. Despite the accepted definition of the Twitter cascade, that is the summation of \textit{Retweet}, \textit{Reply}, and \textit{Quote} activities on a user-generated content, the studies employed the compartmental model to model information propagation on Twitter considered \textit{Retweet} counts only. In this study, we suggest to use the summation of \textit{Retweet}, \textit{Reply} and  \textit{Quote} activities as the components of information propagation on Twitter. Here, we compare SIS and SEIZ compartmental models' performances in predicting the information cascades on Twitter. We also propose a new compartmental model (CD-SEIZ) based on the fact that different type of actions on Twitter require cognitive and physical efforts and processing depths. The details of the proposed method will be explained in the next section.  

\subsection{SIS Model}
This model is one of the simplest compartmental models in which agents travel between two compartments with certain contact rates. The main use of this model in epidemiology is the diseases that people cannot gain immunity and repeatable again, e.g. allergy, flu, etc. In its application to OSN, it allows to model in which the same user can make more than one activity on the same content. Although it has a relatively lower capability to model contagiousness, its simplicity makes it popular in some applications. In this case, users who create a content (\textit{Retweet}, \textit{Reply} or \textit{Quote}) is identified as \textit{Infected} since they viewed and adopted the information. If we suppose that the rate of contact of individuals from \textit{Susceptible} (S) to \textit{Infected} (I) status is $\beta$ and the rate of contact back from \textit{Infected} (I) to  \textit{Susceptible} (S) status is $\lambda$, the following system of ODEs can be generated:

\begin{subequations}
\begin{align}
    \frac{d[S]}{dt}  &= -\beta \frac{S I}{N} + \lambda \frac{I}{N}\\
    \frac{d[I]}{dt}  &= \beta \frac{S I}{N} - \lambda \frac{I}{N}
\end{align}
\end{subequations}
where $N(t)$ denotes the number of individuals in the population which equals to $S(t)+I(t)$.

\subsection{SEIZ Model}
In many compartmental models, as in SIS, individuals (agents) can transition between states directly; however, users' opinions and so their activities may take time since they consider many factors before making a decision. Based on this reasoning, Bettencourt et al. \cite{bettencourt2006power} introduced a new model called SEIZ in which individuals can be in four different states or classes: \textit{Susceptible} (S), \textit{Infected} (I), \textit{Exposed} (E) and \textit{Skeptic} (Z). In the Twitter application, \textit{Susceptible} (S) agents denote users who have not viewed the tweet yet, \textit{Infected} (S) represents the statuses of users who have viewed the tweet and has contributed with one of the actions of \textit{Retweet}, \textit{Reply} or \textit{Quote}. \textit{Exposed} (E) agents are the users who have viewed the tweet but did not take any action yet, but will  \textit{Retweet}, \textit{Reply} or \textit{Quote} after an exposure delay. \textit{Skeptic} (Z) represents the statuses of agents who have viewed the tweet but did not take any action and have selected to ignore it. One of the major development in this model is applying an exposure delay for the state transition of agents from \textit{Susceptible} (S) to \textit{Infected} (S) via another compartment. Although it creates additional complexity, studies demonstrated significant improvement in the prediction of the size of adoption, or OSN cascades \cite{bettencourt2006power,jin2013epidemiological,isea2017new,tipsri2015effect,jin2014misinformation}. With the parameters for rate of contact and transition rates as shown in Figure \ref{fig:SEIZModel} and Table \ref{tab:prm1} and \ref{tab:prm2}, the ODE rules of SEIZ model can be explained as follows: 

\begin{subequations}
\begin{align}
    \frac{d[S]}{dt} &= -\beta \frac{S I}{N} - b \frac{S Z}{N} \\
    \frac{d[E]}{dt} &= -(1-p)\beta \frac{S I}{N} + (1-l)-\beta \frac{ S Z}{N} - \rho \frac{E I}{N} - \epsilon E \\
    \frac{d[I]}{dt} &= p \beta \frac{S I}{N} + \rho \frac{ E I}{N} + \epsilon E\\
    \frac{d[Z]}{dt} &= l b \frac{S Z}{N}
\end{align}
\end{subequations}

\begin{table}[ht]
\label{tab:prm1}
\centering
\caption{Definitions of The Parameters in SEIZ Model}
\begin{tabular}{ll}
 \toprule
\textit{\textbf{Parameter}} & \textit{\textbf{Definition}}\\\hline
 \midrule
$\beta$ & Rate of contact between $S$ and $I$ \\\hline
$b$ & Rate of contact between $S$ and $Z$ \\\hline
$\rho$ & Rate of contact between $E$ and $I$ \\\hline
$p$ & Transition rate of $S$ $\rightarrow$ $I$, given contact with $I$ \\\hline
$1-p$ & Transition rate of $S$ $\rightarrow$ $E$ , given contact with $I$ \\\hline
$l$ & Transition rate of $S$ $\rightarrow$ $Z$, given contact with $Z$ \\\hline
$1-l$ & Transition rate of $S$ $\rightarrow$ $E$ , given contact with $Z$ \\\hline
$\epsilon$ & Incubation rate  \\\hline
\bottomrule
\end{tabular}
\end{table}

\subsection{Proposed Framework (CD-SEIZ Model)}

\begin{figure}[ht]
\centerline{\includegraphics[width=0.6\textwidth]{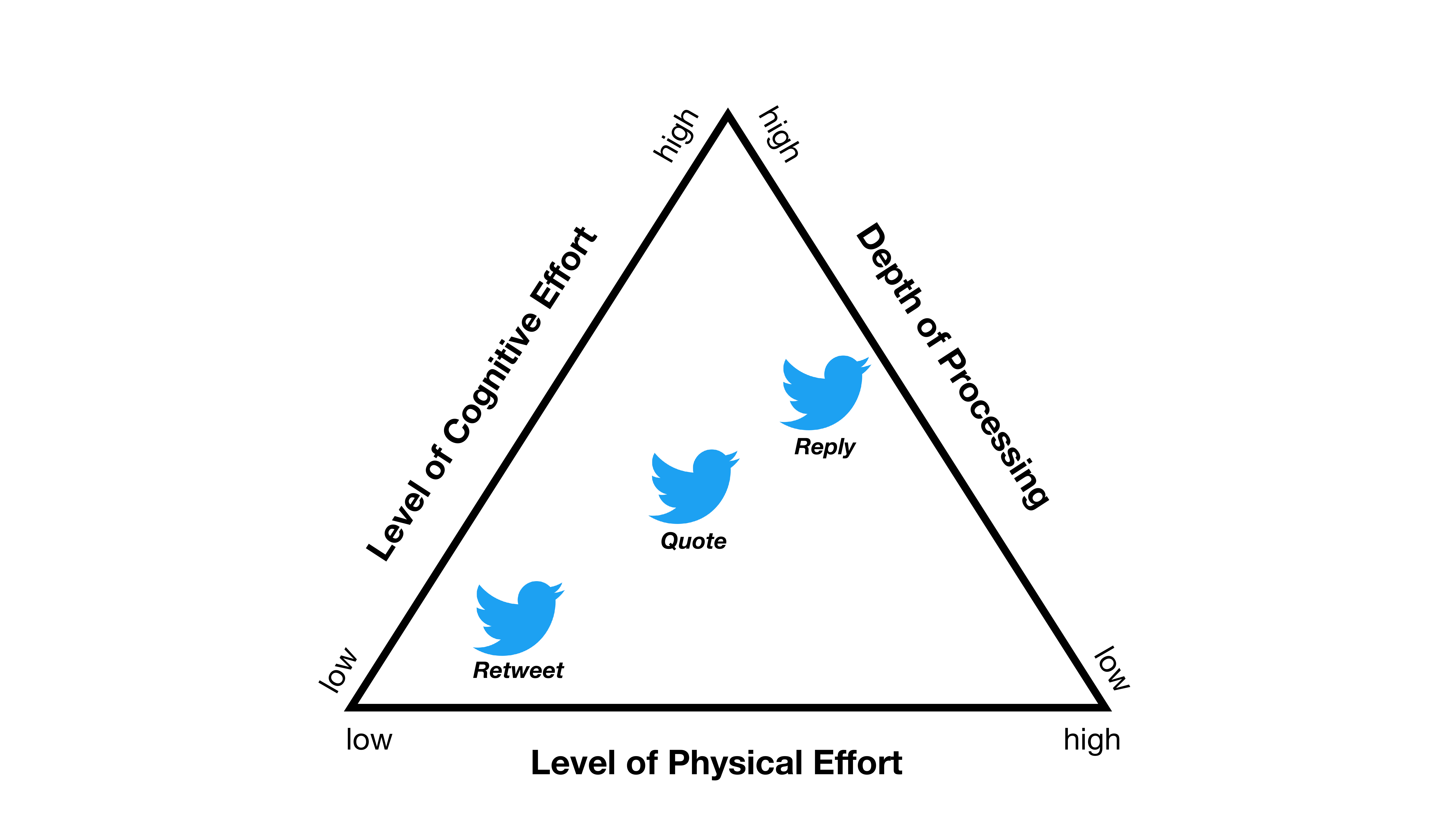}}
\caption{Depth of processing of different twitter activities based on level of cognitive and physical efforts \cite{levens2019using}}
\label{fig:tri}
\end{figure}

\begin{figure}[ht]
\centerline{\includegraphics[width=0.7\textwidth]{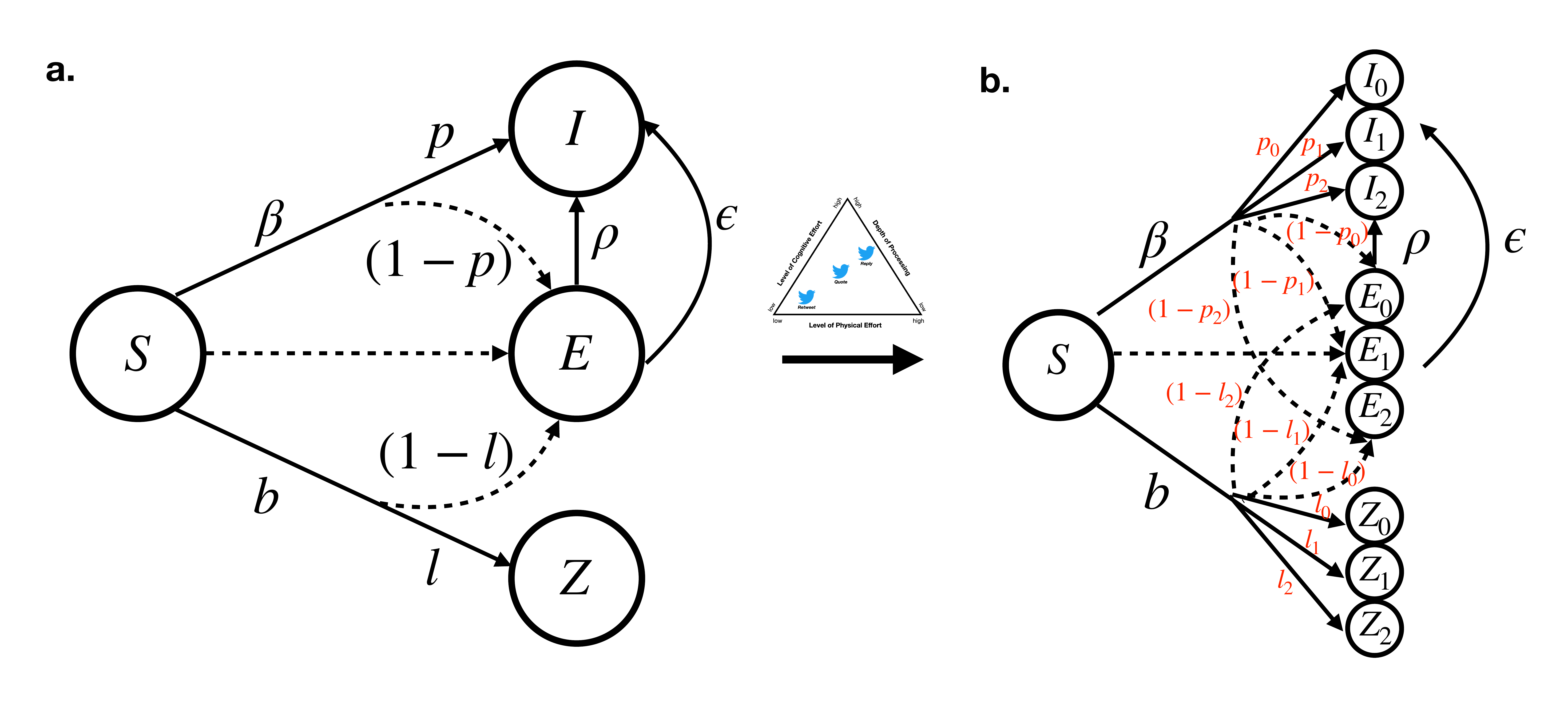}}
\caption{a. Original SEIZ and b. our proposed cognition-driven SEIZ (CD-SEIZ) model. Subscripts 0,1 and 2 refers retweet, quote and reply in a sorted order.}
\label{fig:SEIZModel}       
\end{figure}

Inspired by the current findings \cite{levens2019using}, we argue that different actions in Twitter require different physical and cognitive efforts along with varying processing depth. Levens et al. demonstrated that \textit{Retweet} action requires the least cognitive and physical efforts since it requires an online audience to read the content and push a single button without adding any new content. It is followed by \textit{Quote} action which requires more physical and cognitive effort since users need to process the information and think about writing an original content additionally. \textit{Reply}, on the other hand, has more processing depth than \textit{Quote} because users communicate directly one or more other users in this action. This significant change in level of physical and cognitive efforts, and depth of processing of the different actions on Twitter (Figure \ref{fig:tri}) motivated us to diversify the transition rate between compartments in the SEIZ model for different types of activities. Since the rate of contact of users between compartments depends on the user's engagement on the Twitter platform, we kept all these parameters as constant for each type of activity. The proposed framework of the cognition-driven SEIZ (CD-SEIZ) model and definitions of its parameters can be seen in Figure \ref{fig:SEIZModel} and Table \ref{tab:prm1} and \ref{tab:prm2}, respectively. The ODE rules of the CD-SEIZ model can be written as: 

\begin{subequations}
\begin{align}
    \frac{d[S]}{dt} &= \sum_{i=0}^2{\left [-\beta \frac{S I_i}{N} - b \frac{S Z_i}{N}\right ]}\\
    \frac{d[E_i]}{dt} &= -(1-p_i)\beta \frac{S I_i}{N} + (1-l_i)-\beta \frac{ S Z_i}{N} - \rho \frac{E_i I_i}{N} - \epsilon E_i \\
    \frac{d[I_i]}{dt} &= p_i \beta \frac{S I_i}{N} + \rho \frac{E_i I_i}{N} + \epsilon E_i\\
    \frac{d[Z_i]}{dt} &= l_i b \frac{S Z_i}{N} \quad \text{for $i=\{0,1,2\}$}
\end{align}
\end{subequations}

\begin{table}[ht]
\label{tab:prm2}
\centering
\caption{Definitions of The Parameters in CD-SEIZ Model}
\begin{tabular}{ll}
 \toprule
\textit{\textbf{Parameter}} & \textit{\textbf{Definition}}\\\hline
 \midrule
$\beta$ & Rate of contact between $S$ and $\sum_{i=0}^2{I_i}$ \\\hline
$b$ & Rate of contact between $S$ and $\sum_{i=0}^2{Z_i}$ \\\hline
$\rho$ & Rate of contact between $\sum_{i=0}^2{E_i}$ and $\sum_{i=0}^2{I_i}$ \\\hline
$p_i$ & Transition rate of $S$ $\rightarrow$ $I_i$, given contact with $I_i$ \\\hline
$1-p_i$ & Transition rate of $S$ $\rightarrow$ $E_i$ , given contact with $I_i$ \\\hline
$l_i$ & Transition rate of $S$ $\rightarrow$ $Z_i$, given contact with $Z_i$ \\\hline
$1-l_i$ & Transition rate of $S$ $\rightarrow$ $E_i$ \\\hline
$\epsilon$ & Incubation rate  \\\hline
\bottomrule
\end{tabular}
\end{table}

\subsection{Twitter Data and Cascade Formation}
The data investigated in this work is provided by Leidos Inc\footnote{\url{https://www.leidos.com}} as part of the "Computational Simulation of Online Social Behavior (SocialSim)" DARPA program\footnote{\url{https://www.darpa.mil/program/computational-simulation-of-online-social-behavior}}. The data consists of 1,052,821 tweets related to the disinformation campaigns carried against the White Helmets from April 1st, 2018 to April 30th, 2019. The narratives included within the content of these tweets are mostly attacks against the integrity of the White Helmets' work and mission statement, accusing the organization of being foreign agents, and nullifying the narrative of the chemical attack by censuring the organization of staging the event \cite{starbird2018ecosystem}.

Each tweet in our data set has an identification number (ID) along with its content, date, user identification number (User ID) and the IDs of the tweets that interacted with (retweet/reply/quote). This will help us construct the information cascade associated with our data and trace the argument and discussion that took place in regard to the specific narratives we defined. To build the information propagation cascade, the tweets $t$ ($t=\{t_1, t_2,...,t_n\}$) are separated into two main sets, parent nodes set $P$, and child nodes set $C$. The roots $t_r$ are the tweets that do not have a parent and are the start of the cascades.  Every cascade starts with one root $t_r$ ($ t_r \in P$ $\&$ $t_r \notin C$) and may continue either no message (zero cascade size) or one or many parent-child relations afterwards. Any child of the root that is also a parent of another child is referred as '\textit{parent}' $t_p$, ($ t_p \in P \cap C$), while any child of the root that is not a parent is a called as'\textit{child}' $t_c$ ($ t_c \in C$ $\&$ $t_c \notin P$). Constructing the chain of interaction based on these node sets would give us a temporal tree structure of the conversation regarding a particular narrative.

\begin{figure}[ht]
\begin{center}
\includegraphics[width=0.7\textwidth]{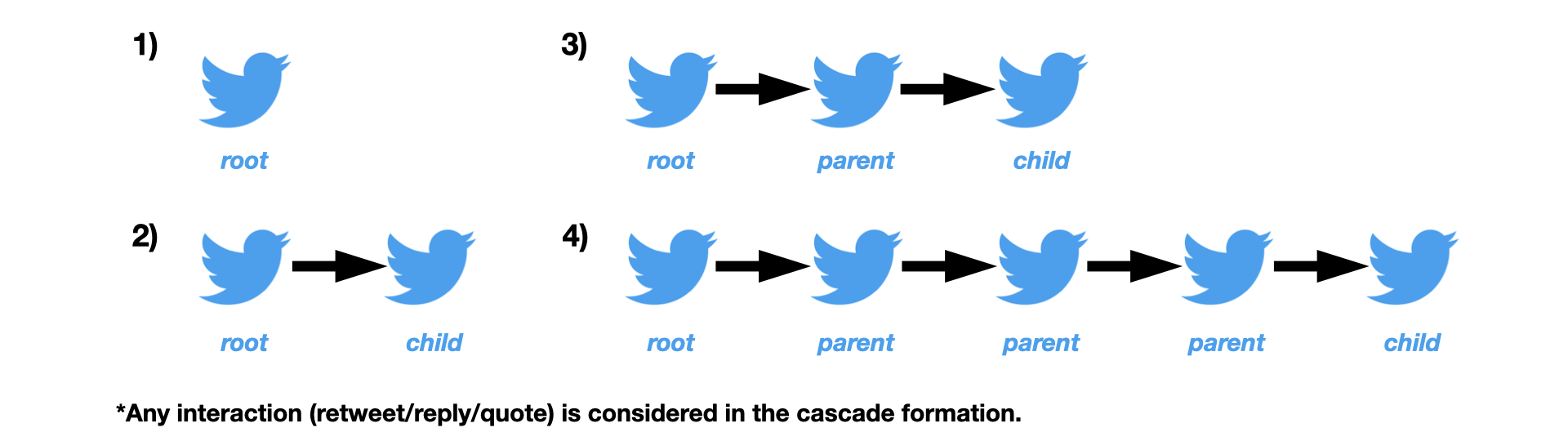}
\caption{Graphical representation of the cascade formation.}
\end{center}
\end{figure}

\section{Results}

\begin{figure}[ht]
\centerline{\includegraphics[width=1.0\textwidth]{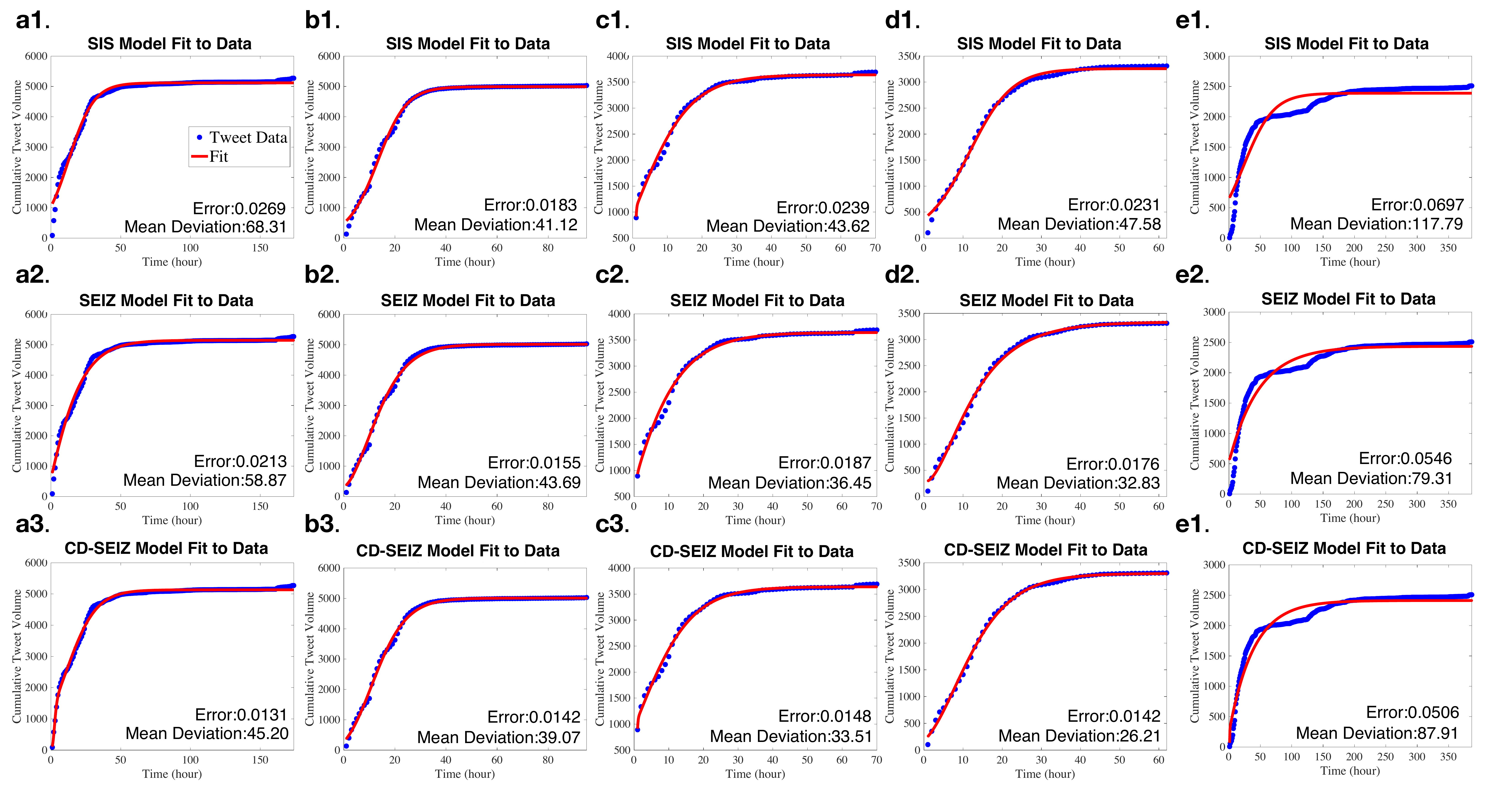}}
\caption{Prediction of the 5 largest information cascades 1.SIS 2.SEIZ and 3.CD-SEIZ model.}
\label{fig:SEIZResults}       
\end{figure}

Figure \ref{fig:SEIZResults} shows the 5 largest Twitter cascade in our data set with blue dots as an hourly time-series. The red plots represent the ODE model fitting of SIS (top), SEIZ (middle) and CD-SEIZ (bottom) models. For a better comparison, the errors and mean deviation values are written in each corresponding figure and calculated as:

\begin{subequations}
\begin{align}
    Error &= \frac{||I(t)-Tweets(t)||_2}{||Tweets(t)||_2}\\
    Mean Deviation &= \sum_{i=1}^n{\frac{|I(t_i)-Tweets(t_i)|}{n}}
\end{align}
\end{subequations}

where $n$ is the number of hours (data points in blue lines). The size of the cascades lower from left (Figure \ref{fig:SEIZResults}a ) to the right (Figure \ref{fig:SEIZResults}e). Although plots for these three models may not seem so distinctive, the error values are more pronounced. In the largest cascades SIS, SEIZ and CD-SEIZ model display 0.0269, 0.0213 and 0.0131 errors respectively. It has been already demonstrated that SEIZ models fit more accurately than SIS model to the Twitter cascade data; however, the CD-SEIZ model reached the best performance with the lowest error. The other Twitter cascades show the same behaviour: SEIZ model is superior than SIS model in each five example since it gives an opportunity to add a delay factor in the transition from \textit{Susceptible} to \textit{Infected} class with a new class of \textit{Exposed}. Furthermore, we assumed that transition rate between states should be different for each type of activity, while rates of contact are same for each. It is reasonable because transition rates are defined with the engagement of users and their network relationship, and users see other user's contents on their feed regardless of the type of the activity; however, the probability (transition rate between states) should differ according to the type of the activity because \textit{Retweet}, \textit{Reply} or  \textit{Quote} activities require different cognitive desire and interest. Results showed that CD-SEIZ model yields better performances in the largest 5 cascades of the data set.

\begin{figure}[ht]
\centerline{\includegraphics[width=0.4\textwidth]{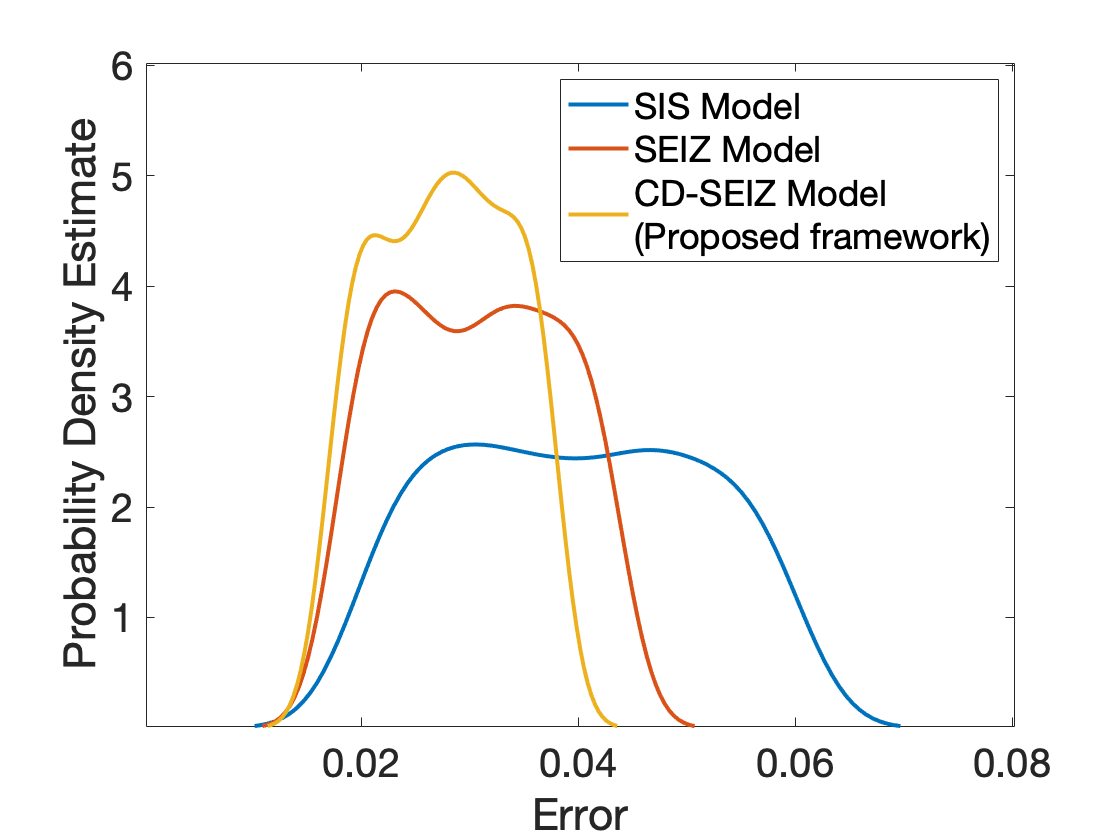}}
\caption{Distribution of the fitting errors in the prediction of 1000 cascades via SIS, SEIZ and CD-SEIZ models}
\label{fig:SEIZResults2}       
\end{figure}

Current studies in the literature generally compared their models with a very limited number of data set since Twitter data collection is a challenging and an expensive process. However, we compared these three models in 1000 cascades in the data set. We did not include more cascades since cascades sizes were getting lower for prediction. The smallest cascade size was 713 in our comparison. Figure \ref{fig:SEIZResults2} demonstrates the distribution of the errors in the prediction of cascade fit via SIS, SEIZ and CD-SEIZ model. Mann-Whitney U test demonstrated the statistical significance of the improvement in CD-SEIZ model compared to SEIZ model with 1.723E-4 p-value.  

\section{Conclusion and Discussion}
The convenient access to OSN platforms, the ability to reach multiple individuals simultaneously, and the free self-expression dynamics within them have motivated researchers to understand the dynamics of the information cascade patterns on these platforms. Previous studies demonstrated the superiority of the SEIZ compartmental model among time series-based studies in the prediction of retweet cascade patterns. On the other hand, quote and reply activities on Twitter play a crucial role in the trajectory of information cascades. Therefore, we considered all retweet, reply, and quote (mention) activities as a considerable reaction. Furthermore, we improved the SEIZ model by varying the transition rates between compartments for these three different activities because it is previously demonstrated that quoting a tweet requires the most physical and cognitive effort, followed by replying and retweeting a tweet, respectively. To test the validity and robustness of our proposed framework, we tested cognition-driven SEIZ (CD-SEIZ) model in 1000 Twitter cascades related to the Syrian White Helmets data set.  In this paper, we have demonstrated how epidemic models can predict true news and rumor stories propagated over Twitter. We have shown that the SEIZ model, in particular, is accurate in capturing the information spread of a variety of news and rumor topics, thereby generating a wealth of valuable parameters to facilitate the analysis of these events. We then demonstrated how these parameters could also be incorporated into a strategy for supporting the identification of Twitter topics as rumor or news. As of now, we are modeling propagation over static data. In the future, we plan to adapt this model for capturing news and rumors in real-time. To the best of our knowledge, we are the first study that employs the SEIZ compartmental model to predict the size of information cascades on Twitter by using retweet and reply and quote activities which are proved to have a significant effect in the diffusion of information. Additionally, we show the comparison between different SIS, SEIZ, and CD-SEIZ models on 1000 different cascades, for the first time, to demonstrate the statistical significance of our results. According to the results, the CD-SEIZ model better fits the information cascade data on Twitter. Further studies might employ the CD-SEIZ model on different data sets to capture the information spread on a variety of topics and compare the parameters of this model in distinguishing true and false news to capture them in real-time.  

\section*{Acknowledgment}
This work was partially supported by grant FA8650-18-C-7823 from the Defense Advanced Research Projects Agency (DARPA).

\bibliographystyle{plain}  
\bibliography{references}
\end{document}